\documentstyle[aps,amssymb,graphicx]{revtex}

\begin{document}
\draft
\title{Parametrization of singularities of the Demia\'{n}ski-Newman spacetimes}
\author{J. Gariel$^1$\thanks{%
Email address: {\tt gariel@ccr.jussieu.fr}}, G. Marcilhacy$^1$, N. O. Santos$%
^{2}$\thanks{%
Email address: {\tt nos@lafex.cbpf.br}} and R. Colistete, Jr.$^1$\thanks{%
Email address: {\tt coliste@ccr.jussieu.fr}}}
\address{$^1$Universit\'e Paris VI, CNRS/UPRESA 7065 Gravitation et
Cosmologie Relativistes \\
Tour 22-12, 4\`eme \'etage, Bo\^{\i}te 142, 4 place Jussieu 75005 Paris,
France.}
\address{$^2$LAFEX, Centro Brasileiro de Pesquisas F\'{\i}sicas \\
Rua Dr. Xavier Sigaud 150, Urca, 22290-180 Rio de Janeiro, RJ - Brazil.}
\date{\today}
\maketitle

\begin{abstract}
We propose a new presentation of the Demia\'{n}ski-Newman (DN) solution of
the axisymmetric Einstein equations. We introduce new dimensionless
parameters $p$, $q$ and $s$, but keeping the Boyer-Lindquist coordinate
transformation used for the Kerr metric in the Ernst method. The family of
DN metrics is studied and it is shown that the main role of $s$ is to
determine the singularities, which we obtain by calculating the Riemann
tensor components and the invariants of curvature. So, $s$ reveals itself as
the parameter of the singular rings on the inner ergosphere.
\end{abstract}


\section{Introduction}

The Kerr metric can be easily obtained from the Ernst equation \cite{Ernst}, 
\begin{equation}
(\xi\xi^{\ast}-1)\nabla^2\xi=2\xi^{\ast}\nabla\xi\cdot\nabla\xi,
\end{equation}
where $\nabla$ denotes the usual three dimensional spatial operator and $\xi$
is a complex potential being function of $\lambda$ and $\mu$ which are the
prolate spheroidal coordinates. By considering, as solution of Eq. (1), 
\begin{equation}
\xi=p_K\lambda+iq_K\mu,
\end{equation}
where $p_K$ and $q_K$ are real constants satisfying 
\begin{equation}
p_K^2+q_K^2=1,
\end{equation}
we can obtain the Kerr solution.

The theorem of Robinson-Carter (see p. 292 of Ref. \cite{Chandra})
demonstrates the uniqueness of this vacuum stationary axisymmetric solution
with an asymptotically flat behavior and smooth convex event horizon
without naked singularity. This solution is characterized by only two
independent parameters, being the mass $M$ and the angular momentum $J=aM$,
where $a$ is angular momentum per unit mass. The Kerr solution has an event
horizon (or outer horizon), a Cauchy horizon (or inner horizon), an outer
ergosphere (or stationary limit surface), an inner ergosphere and a ring
singularity.

In this paper we propose to generalize the solution (2) by considering 
\begin{equation}
\xi =p\lambda +\beta \mu +i(\gamma \lambda +q\mu ),
\end{equation}
where $p,\beta ,\gamma $ and $q$ are real constants. The expression (4) is
also solution of the Ernst equation and corresponds to the
Demia\'{n}ski-Newman (DN) solution \cite{Demianski} as we shall see in
Section II. The DN solution obtained through a complex transformation from
the Kerr solution, namely the introduction of a constant phase factor (see
(7) hereafter), has three independent parameters. The interpretation of the
third parameter, usually denoted $l$, compared to just two parameters in the
Kerr solution, is still not clear. In its place we introduce another
parameter, which we call $s$, which parametrizes the singularities in a
particularly simple way, as we shall see in Section IV. We write the
corresponding metric in Section III.

\section{Choice of the parameters}

The expression (4) is a solution of the equation (1) if the following
conditions are satisfied 
\begin{eqnarray}
p^{2}+\beta ^{2}+\gamma ^{2}+q^{2} &=&1, \\
p\beta &=&-\gamma q.
\end{eqnarray}
Let us recall that the DN solution \cite{Demianski} is usually presented 
\cite{Carmeli} as 
\begin{equation}
\xi =e^{ia_{1}}(p_{DN}\lambda +iq_{DN}\mu ),
\end{equation}
with 
\begin{equation}
p_{DN}^{2}+q_{DN}^{2}=1,
\end{equation}
where $p_{DN},q_{DN}$ and $a_{1}$ are real constants. The comparison between
Eq. (4) and Eq. (7) imposes 
\begin{eqnarray}
p &=&p_{DN}\cos a_{1},\;\;q=q_{DN}\cos a_{1}, \\
\beta &=&-q_{DN}\sin a_{1},\;\;\gamma =p_{DN}\sin a_{1}.
\end{eqnarray}
Then Eqs. (5)--(6) are automatically satisfied, i.e., there is an identity
between Eq. (4) and Eq. (7). In order to write the DN metric in
Boyer-Lindquist (BL) coordinates the following relations are usually
considered (see p. 387 of Ref. \cite{Carmeli}), 
\begin{eqnarray}
\lambda &=&\frac{r-M}{k_{DN}},\;\;\mu =\cos \theta , \\
p_{DN} &=&\frac{k_{DN}}{\sqrt{M^{2}+l^{2}}},\;\;q_{DN}=\frac{a}{\sqrt{%
M^{2}+l^{2}}}, \\
\cos a_{1} &=&\frac{M}{\sqrt{M^{2}+l^{2}}},\;\;\sin a_{1}=\frac{l}{\sqrt{%
M^{2}+l^{2}}},
\end{eqnarray}
with 
\begin{equation}
k_{DN}^{2}=M^{2}+l^{2}-a^{2},
\end{equation}
where $l$ is a third parameter with mass dimension and $r$ and $\theta $ are
spherical coordinates. When $l=0$, the DN metric becomes the Kerr metric.

Here, we shall not proceed in this way and the relations (7)--(14) will not
be used. Instead, we shall present the DN solution as follows.

In place of Eq. (7), we consider the following complex transformation
carried out on the Kerr solution, 
\begin{equation}
\xi =e^{ia_{1}}(p_{K}\lambda +iq_{K}\mu ),
\end{equation}
where the variables $\lambda $ and $\mu $ are the Kerr's ones.

This interpretation of the DN solution, as the Kerr solution with a phase
factor, seems simpler and more natural to us. Indeed Eq. (4) is, for us, a
simple linear extension of the Kerr solution of the Ernst equation.

Then, by identification between Eq. (15) and Eq. (4) we obtain relations
identical to Eqs. (9)--(10), where $p_{DN}$ and $q_{DN}$ are simply replaced
by $p_{K}$ and $q_{K}$ : 
\begin{eqnarray}
p &=&p_{K}\cos a_{1},\;\;q=q_{K}\cos a_{1}, \\
\beta &=&-q_{K}\sin a_{1},\;\;\gamma =p_{K}\sin a_{1}.
\end{eqnarray}

These two Kerr parameters 
\begin{equation}
p_{K}=\frac{k}{M},\;\;q_{K}=\frac{a}{M},
\end{equation}
with 
\begin{equation}
k^{2}=M^{2}-a^{2},
\end{equation}
are linked to the two physical parameters $M$ and $a$. Note that $%
(p_{K},q_{K})$ are different from $(p_{DN},q_{DN})$ given in Eqs. (12),
which are linked to the three parameters $M$, $a$ and $l$, though each
couple obeys the same relation (8) or (3).

To the third parameter $a_{1}$, present in the DN solution (15), we
associate the dimensionless parameter $s$ defined by 
\begin{equation}
s^{2}=\frac{1}{\cos ^{2}a_{1}},\;\;(s\geqslant 1).
\end{equation}

So, the relations (16)--(17) become 
\begin{eqnarray}
p &=&\frac{p_{K}}{s}=\frac{k}{Ms},\;\;q=\frac{q_{K}}{s}=\frac{a}{Ms}, \\
\frac{\gamma }{p} &=&-\frac{\beta }{q}=\sqrt{s^{2}-1}.
\end{eqnarray}
Note that relation (22) also holds from Eqs. (9)--(10), i.e., in the usual
interpretation of the DN solution. With Eq. (22), relation (5) becomes 
\begin{equation}
s^{2}(p^{2}+q^{2})=1,
\end{equation}
which is also in agreement with Eqs. (21).

Finally, instead of Eqs. (11) with relation (14), we introduce BL
coordinates, 
\begin{equation}
\lambda =\frac{r-M}{k},\;\;\mu =\cos \theta .
\end{equation}
where $k$ is defined in relation (19), which is the BL transformation used
for the Kerr solution. In particular, the coordinates are the Kerr's ones,
coherently with the solution (15), which is not the case in the usual DN\
solution for which $\lambda $ depends on $l$ (compare Eqs. (11) with Eqs.
(24)).

\section{Metric coefficients}

The stationary axisymmetric metric, being in the Papapetrou form \cite
{Carmeli,Papapetrou}, reads 
\begin{equation}
ds^{2}=f(dt-\omega d\phi )^{2}-\frac{k^{2}}{f}\left[ e^{2\gamma }(\lambda
^{2}-\mu ^{2})\left( \frac{d\lambda ^{2}}{\lambda ^{2}-1}+\frac{d\mu ^{2}}{%
1-\mu ^{2}}\right) +(\lambda ^{2}-1)(1-\mu ^{2})d\phi ^{2}\right] ,
\end{equation}
where $f,\omega $ and $\gamma $ are functions of $\lambda $ and $\mu $ only.
The real part $P$ and the imaginary part $Q$ of $\xi $, given by the
solution (4), become using relations (22), 
\begin{equation}
P=p\lambda -q\sqrt{s^{2}-1}\mu ,\;\;Q=p\sqrt{s^{2}-1}\lambda +q\mu .
\end{equation}

The Ernst method \cite{Ernst,Carmeli} to determine the metric consists to
make a homographic transformation 
\begin{equation}
\zeta =\frac{\xi -1}{\xi +1},
\end{equation}
where $\zeta $ is of the form 
\begin{equation}
\zeta =f+i\,\Omega ,
\end{equation}
with $f$ being the metric coefficient of the line element (25) and $\Omega$
being the so called twist potential linked to the dragging of spacetime, $%
\omega$, through the differential relations 
\begin{equation}
\omega _{\lambda }^{\prime }=\frac{k(\mu ^{2}-1)}{f^{2}}\Omega _{\mu
}^{\prime },\;\;\omega _{\mu }^{\prime }=\frac{k(\lambda ^{2}-1)}{f^{2}}%
\Omega _{\lambda }^{\prime },
\end{equation}
where indexes indicate to which variable the differentiation, indicated by
primes, is to be taken. Once the partial differential equations (29) are
integrated we obtain $\omega $. Applying this method we have 
\begin{eqnarray}
f &=&\frac{s^{2}[q^{2}(\mu ^{2}-1)+p^{2}(\lambda ^{2}-1)]}{%
s^{2}(p^{2}\lambda ^{2}+q^{2}\mu ^{2})+2[p\lambda -\sqrt{s^{2}-1}q\mu ]+1},
\\
\omega &=&\frac{2k}{s^{2}p}\left[ \frac{q(1+p\lambda )(1-\mu ^{2})+p^{2}%
\sqrt{s^{2}-1}(1-\lambda ^{2})\mu }{q^{2}(\mu ^{2}-1)+p^{2}(\lambda ^{2}-1)}%
\right] .
\end{eqnarray}
From the field equations \cite{Ernst,Carmeli} we also find 
\begin{equation}
e^{2\gamma }=\frac{s^{2}(p^{2}\lambda ^{2}+q^{2}\mu ^{2})-1}{p^{2}(\lambda
^{2}-\mu ^{2})}.
\end{equation}
Substituting the expressions for $p,q$ and $\lambda $, corresponding to Eqs.
(21) and Eq. (24), into Eqs. (30)--(32), we obtain 
\begin{eqnarray}
f &=&\frac{s(r^{2}-2Mr+a^{2}\mu ^{2})}{s(r^{2}+a^{2}\mu
^{2})+2(s-1)M^{2}-2M[(s-1)r+\sqrt{s^{2}-1}a\mu ]}, \\
\omega &=&\frac{2M}{s(r^{2}-2Mr+a^{2}\mu ^{2})}\left\{ a(1-\mu
^{2})[r+(s-1)M]-(r^{2}-2Mr+a^{2})\sqrt{s^{2}-1}\mu \right\} , \\
e^{2\gamma } &=&\frac{s^{2}(r^{2}-2Mr+a^{2}\mu ^{2})}{%
(M-r)^{2}-(M^{2}-a^{2})\mu ^{2}}.
\end{eqnarray}

\section{Singularities of the DN spacetime}

We can write Eq. (33) like 
\begin{equation}
f=\frac{N}{D},
\end{equation}
where 
\begin{eqnarray}
N(r,\mu ) &=&s[r-r_{-}(\mu )][r-r_{+}(\mu )], \\
D(r,\mu ) &=&sr^{2}-2(s-1)Mr+sa^{2}\mu ^{2}-2\sqrt{s^{2}-1}Ma\mu
+2(s-1)M^{2},
\end{eqnarray}
with 
\begin{equation}
r_{-}(\mu )=M-\sqrt{M^{2}-a^{2}\mu ^{2}},\;\;r_{+}(\mu )=M+\sqrt{%
M^{2}-a^{2}\mu ^{2}}.
\end{equation}
The equations $r=r_{-}(\mu )$ and $r=r_{+}(\mu )$, producing $N=0$, define
respectively the inner and outer ergospheres (see p. 316 of Ref. \cite
{Chandra}). The inner horizon or Cauchy horizon, and the outer horizon or
event horizon are given, respectively, by $r_{-}=M-\sqrt{M^{2}-a^{2}}$ and $%
r_{+}=M+\sqrt{M^{2}-a^{2}}$ (see p. 278 of Ref. \cite{Chandra}). $D=0$
defines the singularities of spacetime (see Appendix). In order to have this
second order polynomial equation in $r$ satisfied, we need 
\begin{equation}
r=\frac{1}{s}\left\{ (s-1)M\pm \sqrt{-s^{2}a^{2}(\mu -\mu _{s})^{2}}\right\}
,
\end{equation}
with 
\begin{equation}
\mu _{s}=\sqrt{s^{2}-1}\frac{M}{sa}.
\end{equation}
Hence, to Eq. (40) produce real roots, one needs 
\begin{equation}
\mu =\mu _{s},
\end{equation}
which defines the equation of a cone. In this case $\mu $ is fixed, for a
given $s$, and we can now write $D$ with relation (42) like 
\begin{equation}
D(r)=s(r-r_{s})^{2},
\end{equation}
where 
\begin{equation}
r_{s}=\frac{(s-1)M}{s}.
\end{equation}
Then, according to Eq. (43), the only possible singular points for a given $s
$ are distributed on a sphere with radius 
\begin{equation}
r=r_{s}.
\end{equation}
But at the same time, the singular points have to satisfy condition (42),
and the intersection of the two folds of the cone with the sphere (45)
produce two rings. Hence the singularities for DN spacetime are spread along
two symmetrical rings centered at the $z$ axis. Therefore we can consider
only the upper fold $\theta \in \left[ 0,\pi /2\right] $, remembering that
we have to complete the picture by symmetry. Now, if we eliminate the
parameter $s$ between the two equations (42) and (45), we obtain 
\begin{equation}
r=M-\sqrt{M^{2}-a^{2}\mu ^{2}},
\end{equation}
which we recognize as the equation of the inner ergosphere given just after
Eqs. (39). Hence, each value of $s$ corresponds to a ring singularity, which
is a circle, being the intersection of the inner ergosphere with the cone
(42), see Figure 1. The two circular rings are centered on the $z$ axis at $%
\pm z_{s}$, where 
\begin{equation}
z_{s}=r_{s}\mu _{s}=(s-1)\sqrt{s^{2}-1}\frac{M^{2}}{s^{2}a},
\end{equation}
and the radius $R_{s}$ of the two rings is 
\begin{equation}
R_{s}=r_{s}\sin \theta _{s}=r_{s}\sqrt{1-\mu _{s}^{2}}=\frac{(s-1)M}{s}\sqrt{%
1-\frac{(s^{2}-1)M^{2}}{s^{2}a^{2}}}.
\end{equation}

\begin{figure}
\begin{center}
\includegraphics[scale=0.70,trim=0 70 0 30]{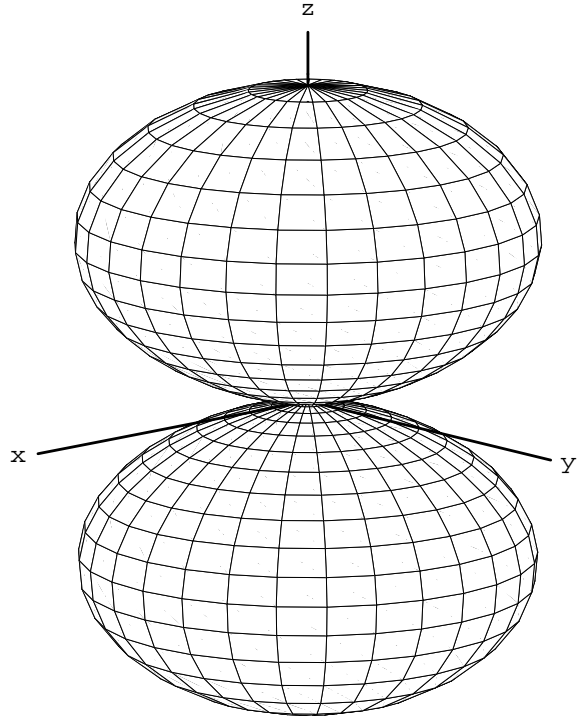}
\caption{
Inner ergosphere of DN spacetime for the following values, $a=4$ and $M=4.2$%
, of the parameters. The vertical axis of revolution is the $z$ axis. The
orthogonal plane $xy$ contains the angle $\phi $, and $\theta $ is the
angle between the $z$ axis and the position vector of a point in the space.
The intersections of the planes $z=cte$ with the inner ergosphere are
the ring singularities. For each value of $s$ there are two
ring singularities, symmetrical with respect to the plane $xy$. For the Kerr
metric $(s=1)$ the ring singularity is reduced to the origin $O$. For
$s_{max}$, the two ring singularities are the poles on the $z$ axis.
}
\end{center}
\end{figure}

We can say that $s$ parametrizes the singular rings, intersections of the
inner ergosphere with a continuous foliation of planes orthogonal to the $z$
axis. For each value of $s$ there is a different spacetime, see Eqs.
(33)--(35). In particular for $s=1$ we have the Kerr spacetime. However the
inner ergosphere is the same, as well as the outer ergosphere and the two
horizons, for all these metrics, i.e., for any $s$ (see Figure 2). Only the
ring singularity changes with $s$, and each ring singularity belongs to the
inner ergosphere.

\begin{figure}
\begin{center}
\includegraphics[scale=0.70,trim=0 -10 0 0]{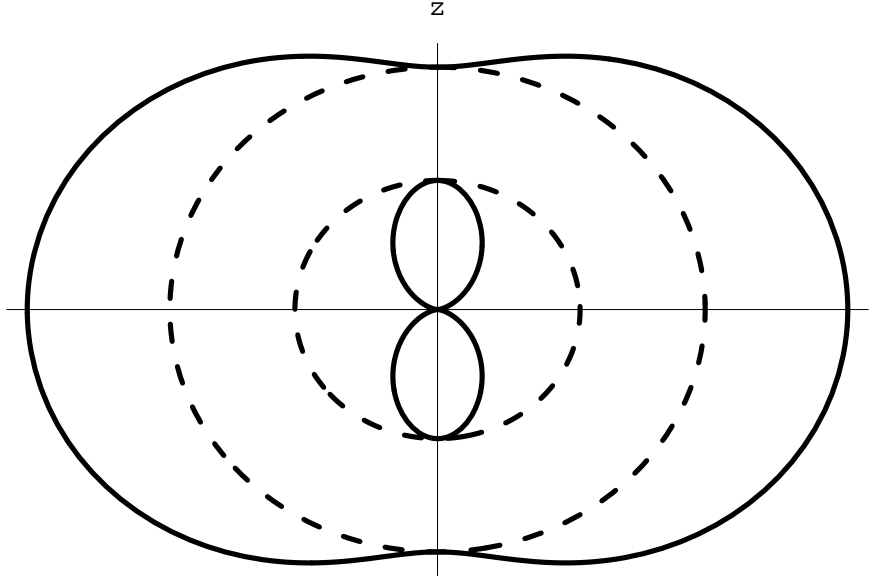}
\caption{
Ergospheres (with solid lines) and horizons (with dashed lines) of DN spacetime
for the following values, $a=4$ and $M=4.2$, of the parameters. We plot here
the intersections of the plane $\phi = cte$ with the surfaces of revolution,
obtained from the rotation of these curves around the $z$ axis.
The radius of Cauchy horizon is here $r_{-} \simeq 2.92$, and the radius
of the event horizon $r_{+} \simeq 5.48$.
}
\end{center}
\end{figure}

By definition $0\leq \mu _{s}\leq 1$, hence from Eq. (41) we have 
\begin{equation}
1\leq s\leq s_{max}=\frac{M}{\sqrt{M^{2}-a^{2}}},
\end{equation}
and from Eq. (44) we have 
\begin{equation}
0\leq r_{s}\leq r_{smax}=r_{-}=M-\sqrt{M^{2}-a^{2}}.
\end{equation}
For $s=1$, which produces Kerr metric, we have from Eq. (44) $r_{s}=0$, or
from Eq. (48) $R_{s}=0$. Hence the Kerr limit minimizes the radius of the
ring singularity and reduces the two ring singularities to just one. The
other metric that minimizes the radius (but produces two symmetrical ring
singularities) happens when 
\begin{equation}
s=s_{max}=\frac{M}{\sqrt{M^{2}-a^{2}}},
\end{equation}
giving, from Eq. (41), $\mu _{s}=1$ and, from Eq. (48), $R_{s}=0$.

We can calculate the maximum radius $R_{smax}$ of the ring singularities
given by the equation (48) for $R_{s}(s)$. Calculating $dR/ds$ we find the
maximum which is for $s$ given by 
\begin{equation}
s=\frac{-1+\sqrt{1+8(1-a^{2}/M^{2})}}{2(1-a^{2}/M^{2})}.
\end{equation}

If $a\ll M$ we see from relation (49) that $1\leq s\leq s_{max}\approx
1+a^{2}/(2M^{2})$, hence up to first order $O(a/M)$ the spacetime reduces to
Kerr spacetime. On the other hand, if $a=0$ and $s\neq 1$ we see that the
spacetime reduces to Taub-NUT spacetime (see p. 387 of Ref. \cite{Carmeli})
and $D(r)=0$ has no real roots for $r$, demonstrating that there are no
singularities in this case.

\begin{figure}
\begin{center}
\includegraphics[scale=0.70,trim=0 70 0 30]{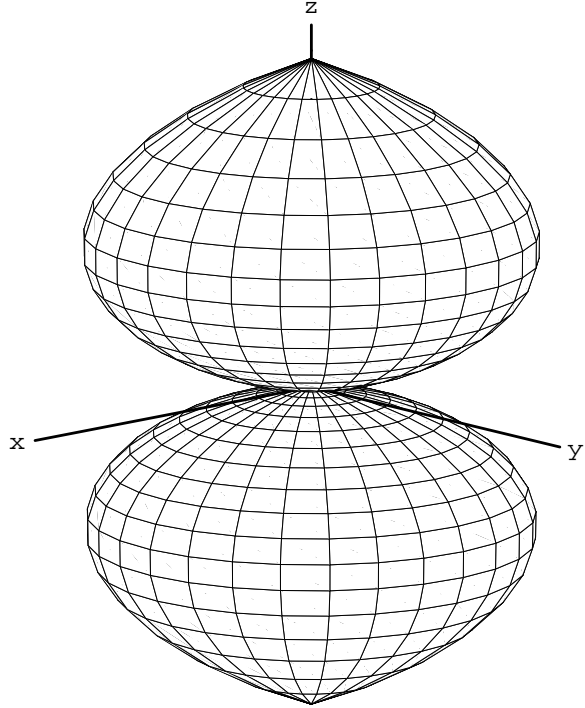}
\caption{
Inner ergosphere of a extreme DN black-hole for the following values, $a=M=4$%
, of the parameters. The intersections of the planes $z=cte$ with the
inner ergosphere are the ring singularities. For each value of $s$
there are two ring singularities, symmetrical with respect to the plane $xy$%
.
}
\end{center}
\end{figure}
\begin{figure}
\begin{center}
\includegraphics[scale=0.70,trim=0 -10 0 0]{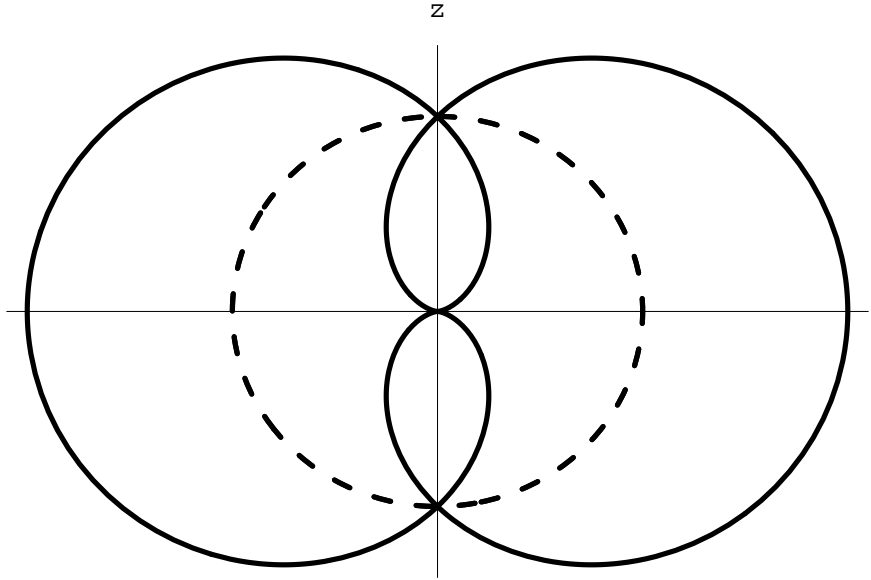}
\caption{
Ergospheres (with solid lines) and horizons (with dashed lines) of a extreme
DN black-hole for the following values, $a=M=4$, of the parameters. We plot
here the intersections of the plane $\phi = cte$ with the surfaces of
revolution, obtained from the rotation of these curves around the
$z$ axis. Notice the continuity of the two ergospheres and of their slopes.
The two horizons are the same, $r_{\pm}=M=4$.
}
\end{center}
\end{figure}

Finally, for the so-called ``extreme black hole'' ($a=M$), equations (41),
(47)--(49) give 
\begin{equation}
\mu _{s}=\frac{\sqrt{s^{2}-1}}{s},\quad z_{s}=\frac{M(s-1)\sqrt{s^{2}-1}}{%
s^{2}},\quad R_{s}=\frac{M(s-1)}{s^{2}},\quad 1\leqslant s<\infty 
\end{equation}
respectively, see Figures 3 and 4. The metric which maximizes the radius of
the ring singularity is obtained from condition (52) for $s=4$, and we
obtain in this case 
\begin{equation}
\mu _{s}=\frac{\sqrt{15}}{4}\ or\ \theta _{s}\simeq 14,48^{o},\quad r_{s}=%
\frac{3M}{4},\quad z_{s}=\frac{3\sqrt{15}M}{16},\quad R_{s}=\frac{3M}{16}.
\end{equation}

\section{Conclusion}

The usual presentation of the DN solution obtained from the Ernst equation
introduces another BL transformation (11) instead of the Kerr's one (24),
and new parameters $(p_{DN},q_{DN})$ which are functions of $(M,a,l)$ and
linked by the relation (8) of Kerr's type.

In our interpretation of the DN solution, we keep the BL coordinate
transformation of Kerr (24), but we introduce new parameters $(p,q)$
depending on $(M,a,s)$ and linked by relation (23) which is no longer of
Kerr's type.

Hence, the DN solution of the field equations for a given source $(a,M)$
constitutes a family of metrics which can be parametrized by a dimensionless
parameter $s$ defined by equation (20). The Kerr solution $(s=1)$ belongs to
this family.

We call generically ``DN black hole'' the set of ergospheres, horizons and
singularities of this family.

Then, the only change that $s$ introduces on the DN black hole structure
concerns the singularities. The ergospheres and horizons are the same for
each metric of the DN family, i.e., whatever $s$, in particular for the Kerr
metric $(s=1)$. $s$ parametrizes only each ring singularity, which always
belongs to the inner ergosphere, including the limiting case of Kerr $(s=1)$.

So the Kerr metric appears as the one which minimizes the ring singularity.

\section*{Appendix}

Here we present the components of $R_{\alpha \beta \gamma \delta }$ for the
metric (25), transformed in spherical coordinates, with relations (33)--(35).
The convention used for the Riemann tensor is $(R^{\alpha }{}_{\beta \gamma
\delta }=-\Gamma ^{\alpha }{}_{\beta \gamma ,\delta }+...)$. Since the
expressions become too long, we restrict to present only the denominators $%
d[R_{\alpha \beta \gamma \delta }]$ of its non null components. We use the
definitions (37)--(38) and $\Delta =r^{2}-2Mr+a^{2}$ producing the following
components : 
\begin{eqnarray}
d[R_{t\phi t\phi }] &=&4sN^{2}D^{3},  \nonumber \\
d[R_{t\phi \theta r}] &=&4\Delta ND^{2},  \nonumber \\
d[R_{t\theta t\theta }] &=&d[R_{t\theta tr}]=d[R_{trtr}]=4\Delta ND^{3}, 
\nonumber \\
d[R_{t\theta \phi \theta }] &=&d[R_{t\theta \phi r}]=d[R_{tr\phi \theta
}]=d[R_{tr\phi r}]=4\Delta N^{2}D^{3},  \nonumber \\
d[R_{\phi \theta \phi \theta }] &=&d[R_{\phi \theta \phi r}]=d[R_{\phi r\phi
r}]=4\Delta N^{3}D^{3},  \nonumber \\
d[R_{\theta r\theta r}] &=&4\Delta D.
\end{eqnarray}
We see that the denominators of all components of the Riemann tensor become
null only if $D=0$, i.e., when the relations (41)--(42), (44)--(45) are
satisfied.

When $s=1$ we reobtain the components of the Riemann tensor for Kerr
spacetime.

The calculation of the invariants of curvature ($R^{2}$, $R^{\alpha \beta
}R_{\alpha \beta }$, $R^{\alpha \beta \gamma \delta }R_{\alpha \beta \gamma
\delta }$, etc)\ confirms that the condition $D=0$ determines the
singularities of DN spacetime.

\section*{Acknowledgments}

R. Colistete Jr. would like to thank CAPES of Brazil for financial support.
N. O. Santos would like to thank the Laboratoire de Gravitation et
Cosmologie Relativistes of Universit\'e Pierre et Marie Curie where part of
this work has been done, for financial aid and hospitality.

\end{document}